\documentclass[11pt]{iopart}
\usepackage{iopams}
\usepackage{xcolor}
\usepackage{graphicx}
\usepackage{geometry}
\makeatletter
\def\footnoterule{\kern-3\p@
  \hrule \@width 2in \kern 2.6\p@} 
\makeatother

\expandafter\let\csname equation*\endcsname\relax

\expandafter\let\csname endequation*\endcsname\relax

\usepackage{amsmath}

\usepackage{scrextend}
\deffootnote{0em}{1.6em}{\thefootnotemark.\enskip}
\usepackage[
	natbib=true,
    style=numeric,
    bibstyle=ieee,
    citestyle=numeric-comp,
    sorting=none]{biblatex}
\usepackage[colorlinks=true]{hyperref}
\definecolor{bluish}{HTML}{228CEE}
\definecolor{greenish}{HTML}{419127}
\definecolor{orangish}{HTML}{FFB347}
\hypersetup{%
  colorlinks=true,
  linkcolor=bluish,
  anchorcolor=bluish,
  citecolor=orangish,
  filecolor=cyan,
  menucolor=magenta,
  runcolor=yellow,
  urlcolor=magenta
}

\begin{document}
\begin{flushright}
{YITP-23-30}
\end{flushright}

\note[Gravitational wave hints black hole remnants as dark matter]{Gravitational wave hints black hole remnants as dark matter}

\author{Guillem Dom\`enech$^{a}$}

\address{$^{a}$Institute for Theoretical Physics, Leibniz University Hannover, Appelstraße 2, 30167 Hannover, Germany.}

\ead{{guillem.domenech}@{itp.uni-hannover.de}}

\author{Misao Sasaki$^{b,c,d}$}
\address{$^{b}$Kavli Institute for the Physics and Mathematics of the Universe (WPI), The University of Tokyo Institutes for Advanced Study, The University of Tokyo, Chiba 277-8583, Japan}
\address{$^{c}$Center for Gravitational Physics, Yukawa Institute for Theoretical Physics, Kyoto University, Kyoto 606-8502, Japan} 
\address{$^{d}$Leung Center for Cosmology and Particle Astrophysics, National Taiwan University, Taipei 10617, Taiwan}

\ead{{misao.sasaki}@{ipmu.jp}}
\vspace{10pt}
\begin{indented}
\item[]\today
\end{indented}

\begin{abstract}
The end state of Hawking evaporation of a black hole is uncertain. Some candidate quantum gravity theories, such as loop quantum gravity and asymptotic safe gravity, 
hint towards Planck sized remnants. If so, the Universe might be filled with remnants of tiny primordial black holes, which formed with mass  $M<10^9\,{\rm g}$. A unique scenario is the case of $M\sim 5\times10^5\,{\rm g}$, where tiny primordial black holes reheat the universe by Hawking evaporation and their remnants dominate the dark matter. Here, we point out that this scenario leads to a cosmological gravitational wave signal at frequencies $\sim 100{\rm Hz}$. Finding such a particular gravitational wave signature with, e.g. the Einstein Telescope, would suggest black hole remnants as dark matter.
\end{abstract}

%
%
%
\maketitle

\section{Introduction}

In 1987, about a decade and a half years after Hawking and Carr \cite{Hawking:1971ei,Carr:1974nx} showed that black holes could form in the early universe, MacGibbon--in a paper published in Nature \cite{MacGibbon:1987my}--already entertained the possibility that cold dark matter could be made of residues left after  black hole evaporation \cite{Hawking:1974rv,Hawking:1975vcx}. The idea that black holes might not evaporate completely already appears by Man'ko and Markov in the proceedings of a singular workshop in Moscow in 1981, called ``Seminar on Quantum Gravity'' \cite{Markov:1984nx} (the proceedings is worth checking). From today's perspective, the possibility of black hole remnants, sometimes also called relics, might not be the most attractive end state of a black hole, albeit it is a plausible one. After all, even if Hawking evaporation leaves something behind, it is not clear how one would directly detect Planck sized objects or prove their existence. Nevertheless, their collective gravitational pull could, e.g., explain a completely invisible dark matter \cite{MacGibbon:1987my} and their bag-of-gold interior might offer a solution to the black hole information loss paradox \cite{Hawking:1976ra}; the latter still under debate \cite{Susskind:1995da}. We refer the reader to ref.~\cite{Chen:2014jwq} for a thorough review on the information loss paradox and black hole remnants.

From a theoretical point of view, remnants might be a consequence of quantum gravity. There is some degree of believe that the theory of quantum gravity would be free from singularities, such as those that appear in the interior of a black hole. Within such candidate quantum gravity theories, a regular black hole might well end up in a stable state after evaporation. 
For instance, regular black hole solutions have been found within loop quantum gravity \cite{Ashtekar:2005cj}, non-commutative geometry \cite{Nicolini:2005vd}, limiting curvature models \cite{Chamseddine:2016ktu,BenAchour:2017ivq} and generalized uncertainty principles \cite{Chen:2002tu}. For more models, see the collection in refs.~\cite{Chen:2014jwq,Hossenfelder:2012jw}. Recently, there is also growing interest within asymptotic safe gravity (see refs.~\cite{Eichhorn:2022bgu,Platania:2023srt} for recent reviews). Note that these are static solutions and it is not clear what would occur when considering the initial collapse of matter.

From a cosmologist perspective, black hole remnants appear to be a ``bonus'' to the rich phenomenology of primordial black holes, or PBHs for short (curious fact: the usage of the acronym PBH dates back to 1975 \cite{Carr:1975qj}, but in lower case letters). The PBH scenario is nowadays a very popular topic, as can be seen from the many recent (and thorough) reviews \cite{Khlopov:2008qy,Sasaki:2018dmp,Green:2020jor,Carr:2020gox,Carr:2020xqk,Escriva:2022duf}. We will be most concerned with tiny ($=M_{\rm PBH}<5\times 10^8\,{\rm g}$) PBHs, since they evaporate much before Big Bang Nucleosynthesis (BBN) \cite{Kawasaki:1999na,Kawasaki:2000en,Hasegawa:2019jsa} and may abundantly leave remnants. Though we will not dwell into the details of the formation of such tiny PBHs, they seem to be easily generated towards the end of inflation by preheating instabilities or quantum stochastic effects \cite{Martin:2019nuw,Animali:2022otk,Briaud:2023eae}. 

Tiny PBHs have an interesting early universe cosmology: they could totally (or partially) reheat the universe \cite{Carr:1976zz,Chapline:1976au,Lidsey:2001nj,Hidalgo:2011fj} and explain the baryon asymmetry of the universe \cite{Turner:1979bt,Turner:1979zj,1976ZhPmR..24...29Z,1976ZhPmR..24..616Z,Barrow:1990he,Dolgov:2000ht,Nagatani:2001nz,Baumann:2007yr,Alexander:2007gj,Fujita:2014hha,Morrison:2018xla,Datta:2020bht,Smyth:2021lkn}. Tiny PBHs will also produce high frequency gravitational waves (GWs) by Hawking evaporation \cite{Dolgov:2000ht,Anantua:2008am,Dolgov:2011cq,Dong:2015yjs,Hooper:2019gtx,Inomata:2020lmk,Masina:2020xhk,Masina:2021zpu,Arbey:2021ysg,Ireland:2023avg}, PBH binaries \cite{Dolgov:2011cq,Inomata:2020lmk} and secondary GWs \cite{Inomata:2020lmk,Papanikolaou:2020qtd,Domenech:2020ssp,Domenech:2021wkk,Domenech:2021ztg,Samanta:2021mdm,Dalianis:2021dbs,Papanikolaou:2022chm}. See ref.~\cite{Inomata:2020lmk,Domenech:2021ztg} for a recollection of GWs associated to tiny PBHs.
For the original works see Carr \cite{Carr:1976zz}, Chapline \cite{Chapline:1976au}, Zeldovich and Starobinsky \cite{1976ZhPmR..24...29Z,1976ZhPmR..24..616Z} and Turner \cite{Turner:1979bt,Turner:1979zj}. On top of all that, PBH remnants\footnote{For a brief recent discussion on how PBHs remnants would not recoil from Hawking evaporation see refs.~\cite{Kovacik:2021qms,Lehmann:2021ijf,DiGennaro:2021vev}.} could account for a fraction or all of the dark matter, if they exist. Any overproduction of remnants would then constrain inflationary models, e.g., see refs.~\cite{Carr:1994ar,Iacconi:2021ltm}.

Very interestingly, it has been recently noticed that some GW products of tiny PBHs might be within the range of future experiments, such as CMB-S4 \cite{Hooper:2019gtx,Masina:2020xhk,Masina:2021zpu,Arbey:2021ysg,Ireland:2023avg} and GW detectors such as ET \cite{Inomata:2020lmk,Papanikolaou:2020qtd,Domenech:2020ssp,Domenech:2021wkk,Domenech:2021ztg,Samanta:2021mdm,Dalianis:2021dbs,Papanikolaou:2022chm} and, perhaps, high frequency GW detectors \cite{Fujita:2014hha,Ireland:2023avg}. In the future, we may find signatures of PBH evaporation in the early universe. In this note, we add a unique signature to the PBH remnant scenario. We show that density fluctuations due to the initial inhomogeneous distribution of PBH leads to (induced) GWs within a fixed frequency range, which enters the LIGO/VIRGO band and could be detected in the future by ET. The rest of the note is organized as follows. We briefly review the PBH remnant scenario as dark matter and the production of low frequency GWs in \S\ref{sec:PBHreview}. We then end with a short discussion in \S\ref{sec:PBHconclusion}. Most of the details of the formulas in this paper can be found, e.g., in refs.~\cite{Carr:1994ar,Baumann:2007yr,Sasaki:2018dmp,Inomata:2020lmk,Domenech:2021ztg}. When needed we use the cosmological parameters of Planck 2018 \cite{Planck:2018vyg}.

\section{PBH remnants and low frequency gravitational waves\label{sec:PBHreview}}

In the tiny PBH scenario we have two basic parameters:  the initial mass of the PBHs at formation $M_{\rm PBH,f}$ and  the initial energy density fraction $\beta=\rho_{\rm PBH,f}/\rho_{\rm total}$ \cite{Sasaki:2018dmp}. For simplicity, we will assume that PBHs form by the collapse of primordial fluctuations with a monochromatic PBH mass function. We discuss later the effects of extended mass functions. Under this assumption, $M_{\rm PBH,f}$ and $\beta$ are related to the cosmic horizon $H$ and the number density of PBHs $n_{\rm PBH}$ at formation.\footnote{
It is also useful to write $M_{\rm PBH,f}$ in grams in terms of $H_{\rm f}$ which gives
\begin{align}\nonumber
M_{\rm PBH,f}\approx 10^{-5} \,{\rm g}\times \frac{M_{\rm pl}}{H_{\rm f}}\,.
\end{align}} For a fixed $M_{\rm PBH,f}$ and $\beta$ we have
\begin{align}\label{eq:Hf}
H_f=4\pi\gamma \frac{M_{\rm pl}^2}{M_{\rm PBH,f}}\approx 5\times 10^{13} \,{\rm GeV}\times \left(\frac{M_{\rm PBH,f}}{1\,{\rm g}}\right)^{-1}\,,
\end{align}
where we used that $\gamma\sim0.2$ \cite{Sasaki:2018dmp} and that $M_{\rm pl}\approx 4.235\times 10^{18}\,{\rm GeV}\approx 4.3\times 10^{-6}\,{\rm g}$. For high energy scale inflation one approximately has $H_{\rm f}\sim 10^{-5}M_{\rm pl}$ and so at least $M_{\rm PBH,f}>1\,{\rm g}$. We also have that\footnote{Since 2022 the International Bureau of Weights and Measures (\href{https://www.bipm.org/en/home}{BIPM}) introduced new prefixes for the metric system. There ``qm'' stands for quectometer defined by ${\rm qm}=10^{-30}\,{\rm m}$, about $10^5$ times larger than the Planck length. Similarly, ``qs'' for quetosecond,  ${\rm qs}=10^{-30}\,{\rm s}$.}
\begin{align}\label{eq:npbh}
n_{\rm PBH,f}=\frac{\rho_{\rm PBH,f}}{M_{\rm PBH,f}}=\frac{3\beta}{4\pi\gamma}H_{\rm f}^3\approx 10^{-3} {\rm qm}^{-3}\times \beta\left(\frac{1\,{\rm g}}{M_{\rm PBH,f}}\right)^{3}\,,
\end{align}
where we used $\rho_{\rm total,f}=3H_f^2M_{\rm pl}^2$ and eq.~\eqref{eq:Hf}.

After formation, tiny PBHs quickly evaporate. The evaporation time reads \cite{Inomata:2020lmk}
\begin{align}\label{eq:teva}
t_{\rm eva}\approx \frac{160\alpha}{3.8\pi g_H(T_{\rm PBH})}\frac{M_{\rm PBH,f}^3}{M_{\rm pl}^4}\approx 400\,{\rm qs}\times \alpha \left(\frac{M_{\rm PBH,f}}{1\,{\rm g}}\right)^{3}\,,
\end{align}
where $g_H(T_{\rm PBH})$ is the spin-weighted degrees of freedom and we introduced the parameter $\alpha$ to take into account the effect of the PBH spin. For no spin $\alpha=0$, while for a near extremal PBH $\alpha\sim1/2$ \cite{Arbey:2019jmj}. In deriving eq.~\eqref{eq:teva} we assumed that the evaporation time is much larger than the formation time and that $g_H(T_{\rm PBH})\approx 108$ \cite{Hooper:2019gtx}.
If we compare the evaporation time \eqref{eq:teva} with the Hubble time at formation, eq.~\eqref{eq:Hf}, 
in a radiation-dominated universe,
\begin{align}\label{eq:tf}
t_{\rm f}\approx\frac{1}{2H_{\rm f}}\approx 10^{-8} \,{\rm qs}\times \left(\frac{M_{\rm PBH,f}}{1\,{\rm g}}\right)\,,
\end{align}
we indeed see that $t_{\rm eva}\gg t_{\rm f}$. 

Many of the discussions that follow will depend on the ratio $t_{\rm eva}/t_f$, which tells us how long these tiny PBHs stayed around. So, let us write it explicitly:
\begin{align}
R_{\rm eva, f}\equiv\frac{t_{\rm eva}}{t_{\rm f}}=\frac{1280\pi\gamma\alpha }{3.8\pi g_H(T_{\rm PBH})}\frac{M_{\rm PBH,f}^2}{M_{\rm pl}^2}\approx 6\times 10^{10}\,\alpha\left(\frac{M_{\rm PBH,f}}{1\,{\rm g}}\right)^{2}\,.
\end{align}
For instance, the condition for the PBH dominance is given by
\begin{align}\label{eq:betamin}
\beta>\beta_{\rm min}=1/\sqrt{R_{\rm eva, f}}\approx  4\times 10^{-6}\alpha^{-1/2}\left(\frac{M_{\rm PBH,f}}{1\,{\rm g}}\right)^{-1}\,.
\end{align}
If $\beta<\beta_{\rm min}$, PBHs never dominate. 
The minimum abundance $\beta_{\rm min}$ is obtained by requiring that $H_{\rm eeq}>H_{\rm eva}$ and using $H_{\rm eeq}/H_{\rm f}\approx\sqrt{2}\beta^2$, where $H_{\rm eeq}$ refers to the Hubble parameter at the time of early radiation-PBH equality after PBH formation. The $\sqrt{2}$ appears after using the exact solutions for a radiation-matter universe (see e.g. eq.~(1.81) in Mukhanov's book \cite{Mukhanov:2005sc}) which is a good approximation since PBH evaporate almost instantaneously \cite{Inomata:2020lmk}. We can also find the amount of the expansion of the universe from PBH formation until PBH evaporation,
\begin{align}\label{eq:aevaaf}
\frac{a_{\rm eva}}{a_{\rm f}}\approx \left\{
\begin{aligned}
&\frac{1}{\sqrt{R_{\rm f,eva}}}\approx  \frac{4\times 10^{-6}}{\alpha^{1/2}}\left(\frac{M_{\rm PBH,f}}{1\,{\rm g}}\right)^{-1};\qquad&\beta<\beta_{\rm min}\,,\\
&\left(\frac{4}{3\beta^{1/2}R_{\rm f,eva}}\right)^{2/3}\approx \frac{8\times10^{-8}}{\beta^{1/3}\alpha^{2/3}}\left(\frac{M_{\rm PBH,f}}{1\,{\rm g}}\right)^{-4/3};\qquad&\beta>\beta_{\rm min}\,,
\end{aligned}
\right.
\end{align}
where, for the case $\beta>\beta_{\rm min}$,
we used the fact that $H=2/(3t)$ in a matter dominated universe. 
We see that larger PBH masses lead to larger number of e-folds. 
The $\beta$ dependence in the case $\beta>\beta_{\rm min}$ can be understood from the fact that the early radiation-PBH equality depends on $\beta$.

We can also compute the temperature of the radiation filling the universe after PBH evaporation. This is given by
\begin{align}\label{eq:Teva}
T_{\rm eva}\approx
2.4 (2.8)\times 10^{10}\,{\rm GeV}\,\alpha^{-1/2}\left(\frac{M_{\rm PBH,f}}{1\,{\rm g}}\right)^{-3/2}\left(\frac{g_{s\star}(T_{\rm eva})}{106.75}\right)^{-1/4}\,,
\end{align}
where the coefficient $2.4(2.8)$ is for the case $\beta<\beta_{\rm min}$($\beta>\beta_{\rm min}$) and $g_{s\star}$ is the effective degrees of freedom for the entropy. Later $g_{\star}$ denotes the effective degrees of freedom for the energy density. To evaluate them we use the fitting formulas provided in ref.~\cite{Saikawa:2018rcs}. Evaporation before BBN, that is $T_{\rm eva}>4\,{\rm MeV}$ \cite{Kawasaki:1999na,Kawasaki:2000en,Hasegawa:2019jsa}, requires $M<5\times 10^8\,{\rm g}$. That is all we need to understand the PBH remnant scenario and the associated low frequency GWs.

\subsection{GWs after PBH domination and evaporation \label{subsec:GWsafter}}

\begin{figure}
\centering
\includegraphics[width=0.9\columnwidth]{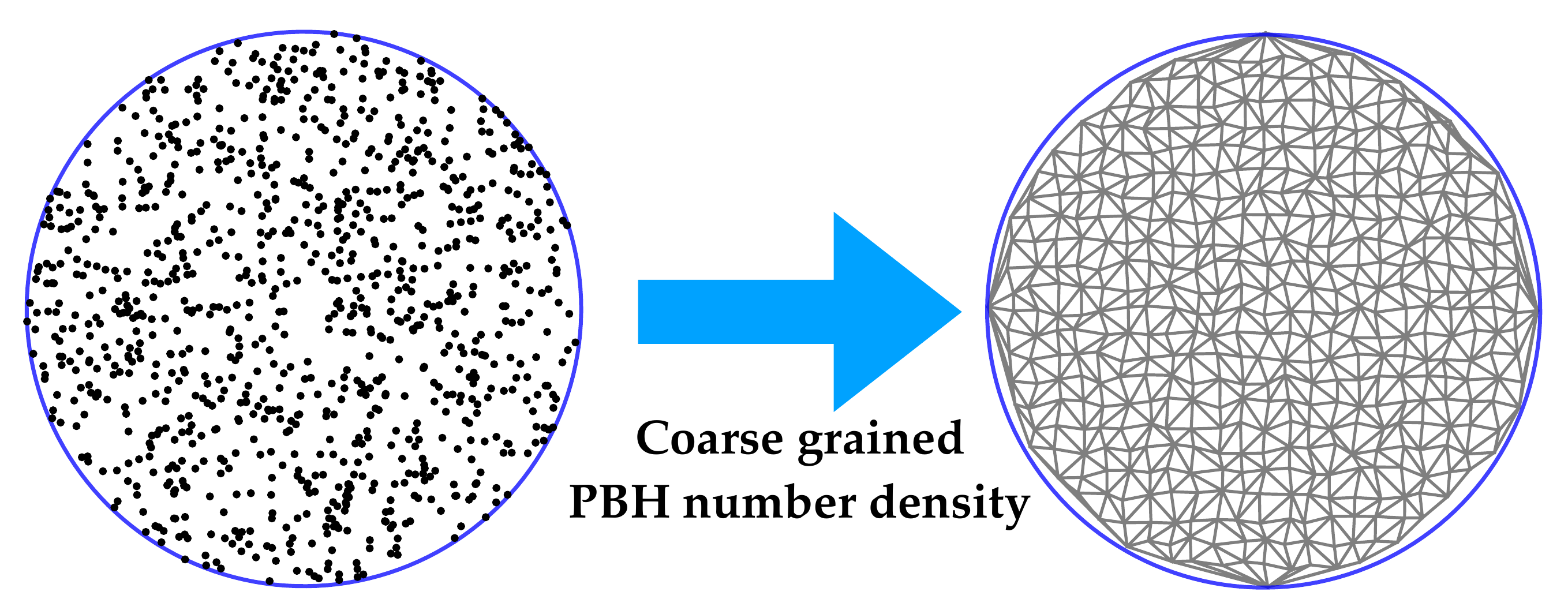}
\caption{PBH form randomly in the early universe approximately according to a uniform distribution. In a coarse grained (fluid) picture this leads to PBH number density fluctuations. These fluctuations are isocurvature in nature because PBH formation leaves a hole in the original radiation fluid. These fluctuations are responsible for a large production of induced GWs. \label{fig:pbhs}}
\end{figure}

PBH formation is rather a rare event. Only those Hubble patches with high enough density contrast will collapse; for Gaussian fluctuations this is exponentially unlikely. Thus, the spatial distribution of PBH formation is to a good approximation uniformly random. This leads to a Poisson type spectrum of PBH number density isocurvature fluctuations \cite{Papanikolaou:2020qtd} (see fig.~\ref{fig:pbhs} for an illustration). Note, however, that the coarse grained, fluid picture for the collection of PBHs breaks down at the mean inter-PBH comoving separation, which is given by
\begin{align}\label{eq:kuv}
k_{\rm uv}\equiv a_{\rm f}\bar d_{\rm f}^{-1}=\left(\frac{4\pi}{3}n_{\rm PBH, f}a_{\rm f}^3\right)^{1/3}= \frac{\beta^{1/3}}{\gamma^{1/3}}a_{\rm f}H_{\rm f}\,.
\end{align}
Equation \eqref{eq:kuv} gives us the high momenta cut-off for the spectrum of number density fluctuations. This is also where fluctuations are larger and the spectrum of (PBH-isocurvature) induced GWs peaks.\footnote{The formation of tiny PBHs from curvature fluctuations also generates induced GWs but those are very high frequency.} We consider from now on, in this subsection, only the case where PBHs dominate the universe, i.e. $\beta>\beta_{\rm min}$. Otherwise, the production of induced GWs is very much suppressed and one needs very large isocurvature fluctuations \cite{Domenech:2021and}.

Curiously, when PBHs dominate, there is a $\beta$ independent relation between $k_{\rm uv}$ and $k_{\rm eva}$, which is given by \cite{Domenech:2020ssp}
\begin{align}\label{eq:kuvkeva}
\frac{k_{\rm uv}}{k_{\rm eva}}=\frac{\beta^{1/3}}{\gamma^{1/3}}\frac{H_{\rm f}}{H_{\rm eva}}\frac{a_{\rm f}}{a_{\rm eva}}=\left( \frac{3R_{\rm eva,f}}{4\gamma}\right)^{1/3}\approx 6000\,\alpha^{1/3}\left(\frac{M_{\rm PBH,f}}{1\,{\rm g}}\right)^{2/3}\,.
\end{align}
This allows us to easily find the peak frequency of the GWs today by using that
\begin{align}
f_{\rm eva}=\frac{k_{\rm eva}}{2\pi a_0}&\approx 3\times 10^{-5}{\rm Hz}\left(\frac{T_{\rm eva}}{1 \rm TeV}\right)\left(\frac{g_{\star}(T_{\rm eva})}{106.75}\right)^{1/2}\left(\frac{g_{s\star}(T_{\rm eva})}{106.75}\right)^{-1/3}\nonumber\\&
\approx 734 \, {\rm Hz}\,\alpha^{-1/2}\left(\frac{M_{\rm PBH,f}}{1\,{\rm g}}\right)^{-3/2}\left(\frac{g_{\star}(T_{\rm eva})}{106.75}\right)^{1/4}\left(\frac{g_{s\star}(T_{\rm eva})}{106.75}\right)^{-1/3}\,,
\end{align}
which yields
\begin{align}
f_{\rm uv}
\approx 4.4\times 10^{6}\, {\rm Hz}\,\alpha^{-1/6}\left(\frac{M_{\rm PBH,f}}{1\,{\rm g}}\right)^{-5/6}\left(\frac{g_{\star}(T_{\rm eva})}{106.75}\right)^{1/4}\left(\frac{g_{s\star}(T_{\rm eva})}{106.75}\right)^{-1/3}\,.
\end{align}

The calculation on the dominant contribution to the amplitude of GWs can be found in ref.~\cite{Domenech:2020ssp} (see also \cite{Papanikolaou:2020qtd,Inomata:2019ivs,Inomata:2019zqy}). Here we only sketch the production of induced GWs and their amplitude after a PBH dominated universe. Let us first give the result and then explain it. The induced GW spectrum from a PBH dominated universe evaluated today is given by
\begin{align}
\Omega_{\rm GW,0}h^2=1.62\times 10^{-5}\,c_g\,\Omega^{\rm peak}_{\rm GW,eva}\left(\frac{k}{k_{\rm uv}}\right)^{11/3}\Theta(k-k_{\rm uv})\,,
\end{align}
where
\begin{align}
c_g\equiv\left(\frac{\Omega_{\rm r,0}h^2}{4.18\times 10^{-5}}\right)\left(\frac{g_{\star}(T_{\rm eva})}{106.75}\right)\left(\frac{g_{s\star}(T_{\rm eva})}{106.75}\right)^{-4/3}\,,
\end{align}
which takes into account the redshift of the GW energy density from evaporation until today as well as the change of relativistic degrees of freedom, and
\begin{align}\label{eq:Omegapeak}
  \Omega^{\rm peak}_{\rm GW,eva}\approx \frac{\beta ^{16/3} \gamma ^{8/3} }{1536\,\times {2}^{1/3} \sqrt{3} \pi }\left(\frac{k_{\rm uv}}{k_{\rm eva}}\right)^{17/3}\approx\frac{\alpha^{17/9}}{3} \left(\frac{\beta}{10^{-3}}\right)^{16/3}\left(\frac{M_{\rm PBH,f}}{1\,{\rm g}}\right)^{34/9}\,,
\end{align}
which gives the peak amplitude right after PBH evaporation. 

Now, let us roughly explain the origin of the amplitude eq.~\eqref{eq:Omegapeak}. First of all, the factor $\beta^{16/3}$ is the suppression due to the fact that PBHs are formed during the radiation era and the amplitude of curvature fluctuations decay until PBHs dominate the universe. The suppression goes as $\sim (k_{\rm eeq}/k_{\rm uv})^2\propto \beta^{4/3}$, where $k_{\rm eeq}=a_{\rm eeq}H_{\rm eeq}$ is the mode that enters the horizon at the early radiation-PBH equality. Induced GWs, as a secondary effect, are proportional to the four-point function of curvature fluctuations. Hence, we obtain $(\beta^{4/3})^4=\beta^{16/3}$. The factor involving $k_{\rm uv}/k_{\rm eva}$ is more interesting. During PBH domination, the curvature perturbation is constant on all scales and density fluctuations grow proportional to the scale factor. On scales corresponding to $k_{\rm uv}$, the PBH number density fluctuations at evaporation have grown by a factor $(k_{\rm uv}/k_{\rm eva})^2$, which is very large. And then, PBHs almost suddenly evaporate: huge pressureless density fluctuations are converted into huge radiation fluctuations which generates a huge wake in the velocities of the radiation fluid. For the induced GW spectrum, which is proportional to four gradients of velocities, we get $((k_{\rm uv}/k_{\rm eva})^2)^4=(k_{\rm uv}/k_{\rm eva})^8$. To go from $8$ to $17/3$ one needs to account that only a very narrow window of scalar modes contribute to the integral, which is proportional to $k_{\rm eva}/k_{\rm uv}$, this gets us to $7$. The rest comes from the ``almost'' sudden evaporation \cite{Inomata:2020lmk}: very short wavelength modes actually feel the time dependence of Hawking evaporation, which goes as $M_{PBH}(t)\sim (1-t/t_{\rm eva})^{1/3}$, and suppresses curvature fluctuations by a factor $(k_{\rm eva}/k_{\rm eva})^{1/3}$. Thus, we have $8-1-4\times1/3=17/3$. We can then use eq.~\eqref{eq:Omegapeak} to place upper bounds on $\beta$ from BBN constraints, namely \cite{Domenech:2020ssp}
\begin{align}\label{eq:betamax}
\beta < 10^{-3}\left(\frac{M_{\rm PBH,f}}{1\,{\rm g}}\right)^{-17/24}\,.
\end{align}

It is important to emphasize that eq.~\eqref{eq:Omegapeak} should be understood as a rough order of magnitude estimate, because at some point during PBH domination, the number density fluctuations exceed unity. Furthermore, the amplitude is very sensitive to the width of the PBH mass function. For a log-normal with logarithmic width $\sigma\sim 1$ the amplification is negligible \cite{Inomata:2020lmk}. In any case, let us note that for the parameters of interest, the curvature perturbation and its time derivative are always smaller than unity and well within the perturbative regime.

\subsection{PBH remnants as dark matter}

Let us collect all the previous results and assume that PBH evaporation leaves behind Planck relics with mass
\begin{align}
m_{\rm relic}=r M_{\rm pl}\,,
\end{align}
where $r>1$ is a free parameter. 
The fact that evaporation stops when $M_{\rm PBH}(t_{\rm end})=m_{\rm relic}$ does not affect the calculations in \S\ref{subsec:GWsafter}. That is, $t_{\rm end}=t_{\rm eva}-\delta t$,
where $\delta t/t_{\rm eva}=O(t_{\rm eva}M_{\rm pl})$,
which is negligibly tiny.
Thus, let us use the previous results and require that PBH remnants occupy a fraction $f_{\rm relic}$ of the total dark matter (DM) today, that is
\begin{align}
f_{\rm relic}\equiv\frac{\rho_{\rm relic}}{\rho_{\rm DM}}\,.
\end{align}

\begin{figure}
\centering
\includegraphics[width=0.6\columnwidth]{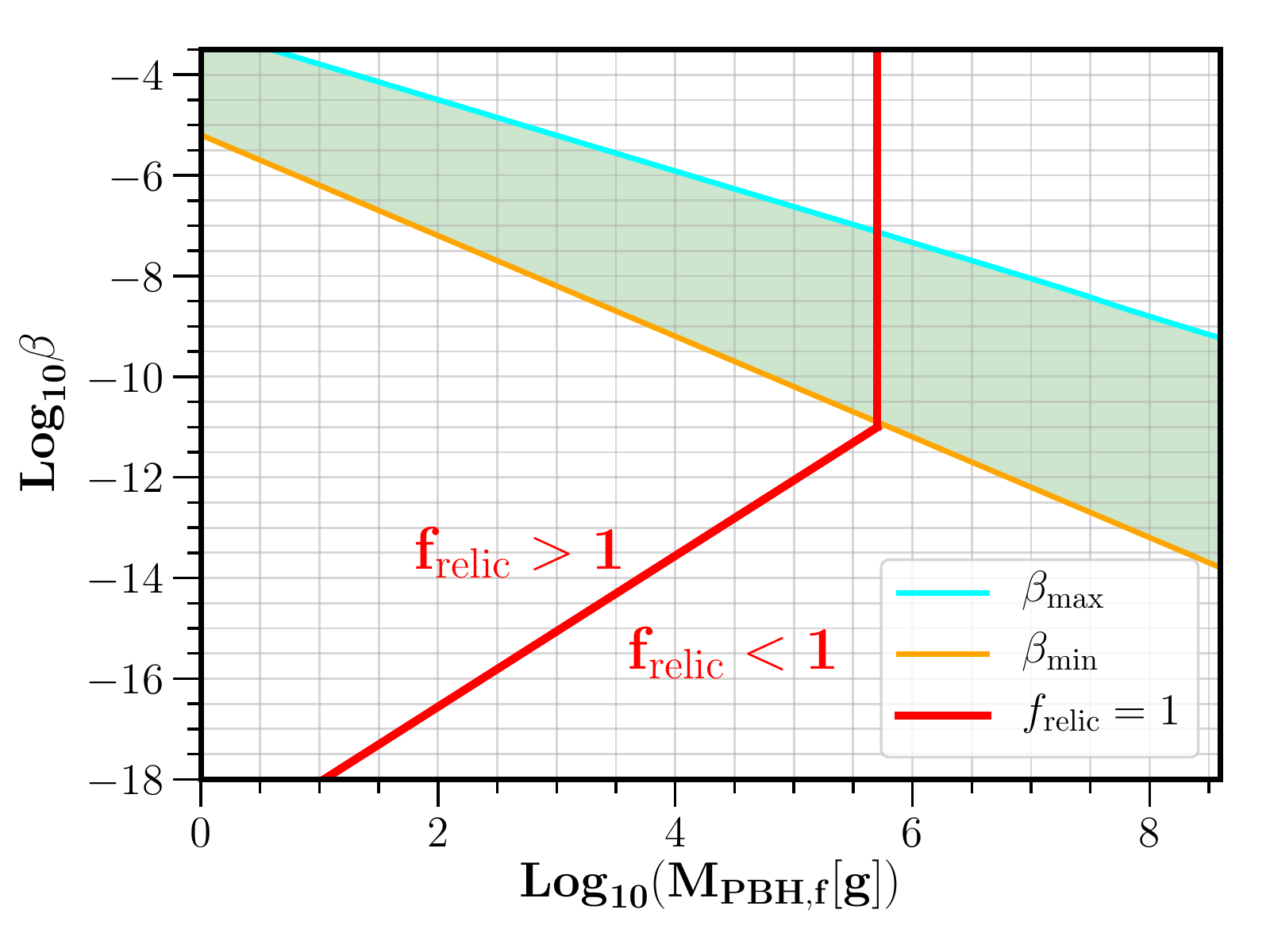}
\caption{Parameter space in terms of the two basic parameters of the model, $\beta$ and $M_{\rm PBH,f}$. The orange line shows the minimum value of $\beta$ \eqref{eq:betamin} to have PBH domination. The cyan line comes from requiring that induced GWs are not overproduced at BBN \eqref{eq:betamax}. The shaded region show the allowed parameter space to have PBH reheating. The red line shows the required values so that PBH remnants are the total dark matter \eqref{eq:mpbhfrelic}. \label{fig:beta}}
\end{figure}

Extrapolating backwards from today until evaporation, using that $\rho_{\rm DM}\propto a^{-3}$, we have that
\begin{align}
\Omega_{\rm relic}\big|_{\rm eva}&\approx7\times 10^{-13}f_{\rm relic}\left(\frac{T_{\rm eva}}{1\rm TeV}\right)^{-1}\left(\frac{g_{s\star}(T_{\rm eva})}{106.75}\right)\left(\frac{g_{\star}(T_{\rm eva})}{106.75}\right)^{-1}\nonumber\\&
\approx 2.9 \,(2.5)\times 10^{-20}f_{\rm relic}\sqrt{\alpha}\left(\frac{M_{\rm PBH,f}}{1\rm g}\right)^{-3/2}\left(\frac{g_{s\star}(T_{\rm eva})}{106.75}\right)\left(\frac{g_{\star}(T_{\rm eva})}{106.75}\right)^{-3/4}\,,
\end{align}
where again the value between brackets is for $(\beta>\beta_{\rm min})$ and we used that $\rho_{\rm DM,eq}=3 H_{\rm eq}^2M_{\rm pl}^2/2$. If we now extrapolate forwards from the PBH formation, we have that at evaporation
\begin{align}
\Omega_{\rm relic}\big|_{\rm eva}
=\frac{\beta m_{\rm relic}}{M_{\rm PBH,f}}\frac{H_{\rm f}^2}{H_{\rm eva}^2}\frac{a_{\rm f}^3}{a_{\rm eva}^3}
\approx\left\{
\begin{aligned}
& \dfrac{m_{\rm relic}}{M_{\rm PBH,f}}{\beta}\sqrt{R_{\rm eva,f}}\approx r\beta\sqrt{\alpha}&(\beta<\beta_{\rm min})\\
& \dfrac{m_{\rm relic}}{M_{\rm PBH,f}}&(\beta>\beta_{\rm min})
\end{aligned}
\right.\,.
\end{align}
In this way, we can draw the parameter space where PBH remnants can be a fraction $f_{\rm relic}$ of dark matter. In general we find that for a fixed $f_{\rm relic}$,
\begin{align}\label{eq:mpbhfrelic}
{M_{\rm PBH,f}}\approx10^6{\rm g}\left\{
\begin{aligned}
&  2.4\dfrac{r^{2/3}}{f^{2/3}_{\rm relic}}\left(\frac{\beta}{10^{-10}}\right)^{2/3}\left(\frac{g_{s\star}(T_{\rm eva})}{106.75}\right)^{-2/3}\left(\frac{g_{\star}(T_{\rm eva})}{106.75}\right)^{1/2}&(\beta<\beta_{\rm min})\\
& 0.5\dfrac{r^{2/5}}{f^{2/5}_{\rm relic}\alpha^{1/5}} \left(\frac{g_{s\star}(T_{\rm eva})}{106.75}\right)^{-2/5}\left(\frac{g_{\star}(T_{\rm eva})}{106.75}\right)^{3/10} &(\beta>\beta_{\rm min})
\end{aligned}
\right.\,.
\end{align}
We show the parameter space in fig.~\ref{fig:beta}.

Note that, if there was a PBH dominated stage in the early universe ($\beta>\beta_{\rm min}$), and if the PBH remnants totally account for the dark matter ($f_{\rm relic}=1$), the initial PBH mass is uniquely determine as
\begin{align}\label{eq:predictionmpbh}
M_{\rm PBH,f}\approx5 \times 10^{5}{\rm g}\times \frac{r^{2/5}}{\alpha^{1/5}}\,.
\end{align}
This correspond to an evaporation temperature of $T_{\rm eva}\approx80\,{\rm GeV}$. Note that value is in agreement with ref.~\cite{Baumann:2007yr}.
This case corresponds to an induced GW signal with peak at
\begin{align}\label{eq:predictionf}
f_{\rm uv}\approx 80\,{\rm Hz}\times \alpha^{-1/6}\,.
\end{align}
The frequency \eqref{eq:predictionf} corresponds to an inter-PBH comoving separation of $600\,{\rm km}$, which is also the mean separation between the remnants.

\begin{figure}
\centering
\includegraphics[width=0.6\columnwidth]{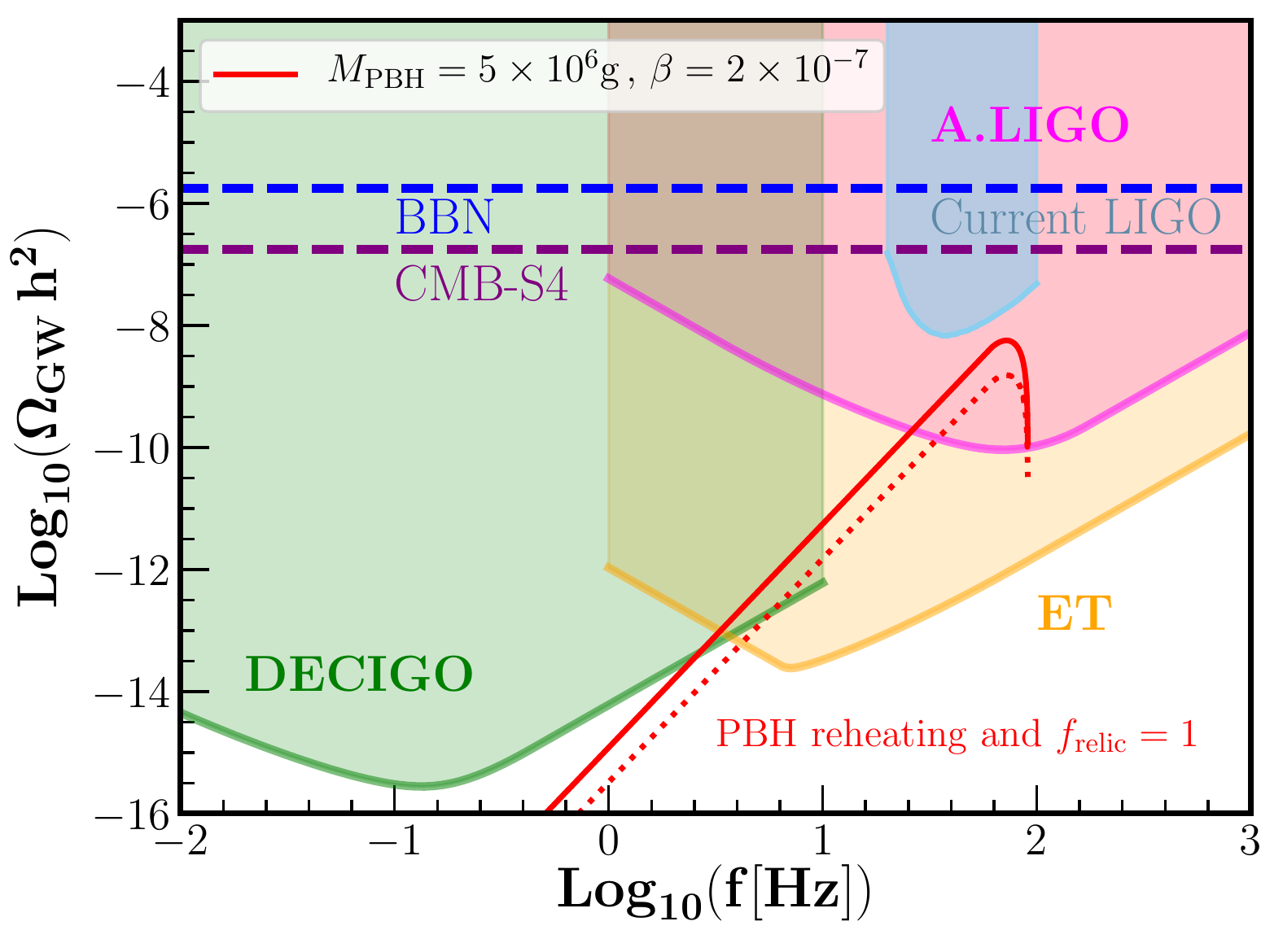}
\caption{Spectral density of GWs induced by PBH number density fluctuations after PBH evaporation. The solid red line corresponds to the signal from the PBH reheating plus dark matter remnants scenario with $M_{\rm PBH,f} =5\times 10^5$~g and $\beta=2\times 10^{-7}$. The dahsed red line is the same signal but for an almost extremal Kerr black hole, i.e. $\alpha=1/2$. Only the amplitude of the signal depends on $\beta$. The peak frequency is solely determined by the PBH mass which is fixed in this scenario. We also show for illustration the power-law integrated sensitivity curves for DECIGO, ET and Advanced LIGO experiments~\cite{Thrane:2013oya}. In light blue we plot theupper bounds on the GW background from the LIGO/VIRGO/KAGRA collaboration \cite{KAGRA:2021kbb}.\label{fig:gws}}
\end{figure}

The PBH reheating scenario with PBH remnants has a unique prediction for the peak frequency of the induced GW spectrum. For a fixed PBH mass, the amplitude of the GW spectrum \eqref{eq:Omegapeak} only depends on $\beta$ and its value today is given by 
\begin{align}\label{eq:Omegapeaktoday}
  \Omega^{\rm peak}_{\rm GW,0}h^2\approx4\times 10^{-11}r^{68/45}\alpha^{17/15}\left(\frac{\beta}{10^{-8}}\right)^{16/3} \,.
\end{align}
We plot the induced GW spectrum in fig.~\ref{fig:gws}. Recall that $1>\alpha>1/2$ and $r\sim O(1)$ so that the predictions \eqref{eq:predictionmpbh}, \eqref{eq:predictionf} and \eqref{eq:Omegapeaktoday} do not depend much on whether PBHs have spin or if the remnant is a bit larger than the Planck length. 
Although this signal could also be present without remnants, detecting a peak right at this frequency would be a strong indication that the PBH remnants is the dark matter.

\section{Conclusions\label{sec:PBHconclusion}}

The existence of remnants after Hawking evaporation is suggested in some theories of quantum gravity \cite{Ashtekar:2005cj,Eichhorn:2022bgu,Platania:2023srt}. 
The remnants could play an important role in the information loss paradox and in cosmology \cite{Chen:2014jwq}. 
Here we focused on the possibility that the universe is filled with the remnants of tiny PBHs which evaporated well before Big-Bang nucleaosynthesis. For some parameter space the PBH remnants could account for all the dark matter and reheat the universe \cite{MacGibbon:1987my,Chen:2002tu,Baumann:2007yr}.

One of the problems of the PBH reheating plus dark matter remnants scenario was that it seemed almost impossible to probe. 
In this note, we pointed out that it may lead to a unique prediction for the GW background: a peaked signal at a frequency $\sim80$\,Hz, close to the peak sensitivity of LIGO/VIRGO/KAGRA, ET and the cosmic explorer. While this is not definitive evidence of remnants as dark matter, finding a peak at such a precise frequency would give a strong indication of the PBH reheating plus remnants scenario. This could be further probed by additional signatures of high frequency GWs and the effective number of species \cite{Fujita:2014hha,Hooper:2020evu,Masina:2020xhk,Arbey:2021ysg,Ireland:2023avg}. 

A remaining issue is to derive a more accurate estimate for the amplitude of (PBH isocurvature) induced GWs, since the PBH number density fluctuations reach the nonlinear regime close to the final stage of evaporation \cite{Inomata:2020lmk,Domenech:2020ssp}. However, this requires sophisticated numerical simulations. 
Another issue is the effect of a finite width in the PBH mass function. 
While these issues might reduce the amplitude of the induced GW spectrum, the peak frequency would not be significantly affected.
Thus the prediction for a GW background peaked at $\sim 80$\,Hz seems robust in the scenario of PBH remnants as dark matter within $O(1)$ factors. 

\section*{Acknowledgments}

GD thanks A.~D.~Rojas, C.~D.~Rojas and D.~Rojas for their support and A.~Platania for useful correspondence on black holes in asymptotic safe gravity. GD is supported by the DFG under the Emmy-Noether program grant no. DO 2574/1-1, project number 496592360. Kavli IPMU is supported by World Premier International Research Center Initiative (WPI), MEXT, Japan. This work was supported in part by the JSPS KAKENHI grant Nos. 19H01895, 20H04727, and 20H05853.

\printbibliography

\end{document}